\documentclass[11pt,twoside]{article}

\usepackage{asp2006}
\usepackage{epsf}
\usepackage{lscape}

\begin{document}
\title{Infrared spectral energy distributions of submillimetre galaxies} 
\author{Alexandra Pope$^{1}$, Ranga Chary$^{2}$, Mark Dickinson$^{3}$, Douglas Scott$^{1}$}   
\affil{$^{1}$University of British Columbia, Vancouver, BC, Canada \\
$^{2}$Spitzer Science Center, Pasadena, CA, USA \\
$^{3}$National Optical Astronomy Observatory, Tucson, AZ, USA
}    

\begin{abstract}
Submillimetre (sub-mm) galaxies have very high infrared (IR) luminosities and are thousands of times more numerous at $z\sim2$ than local ultra-luminous IR galaxies. They therefore represent a key phase in galaxy evolution which can be missed in optical surveys. Determining their contribution to the global star formation rate requires dissecting their IR emission into contributions from starbursts (SB) and active galactic nuclei (AGN). There are several examples of AGN systems which masquerade as SBs in either the IR or X-ray, and SBs can often look like AGN in some wavebands. 
A combination of SB and AGN emission is not unreasonable, given models of merger-driven evolution. To assess in detail what powers the intense IR luminosity of sub-mm galaxies it is important to obtain a complete multi-wavelength picture.
Mid-IR spectroscopy is a particularly good probe of where the intense IR luminosity is coming from. We present the first results from a program to obtain {\it Spitzer} IRS spectroscopy of a sample of high redshift galaxies in the GOODS-N field, a large fraction of which are sub-mm galaxies. This field is already home to the deepest X-ray, optical, IR and radio data. We piece together the sub-mm data with the {\it Spitzer} photometry and IRS spectra to provide a well sampled IR spectral energy distribution (SED) of sub-mm galaxies and determine the contribution to the bolometric luminosity from the AGN and SB components. 
\end{abstract}

\section{Introduction}
Enormous effort has gone into understanding the nature of sub-mm selected galaxies since they were first discovered with the Submillimetre Common User Bolometer Array (SCUBA, Holland et al.~1999) just under a decade ago (e.g.~Smail et al.~1997; Barger et al.~1998; Hughes et al.~1998). A general understanding of the sub-mm population and its role in galaxy evolution is severely limited by the poor spatial resolution of current sub-mm telescopes and the faintness of these sources at other wavelengths (e.g.~Pope et al.~2005). The sub-mm population appears to consist of massive objects (Borys et al.~2005; Greve et al.~2005) at $z\sim2-3$ (Chapman et al.~2005; Pope et al.~2005). These galaxies are rare and have very high star formation rates (SFRs, Lilly et al.~1999; Scott et al.~2002), and hence may represent an early phase in the evolution of massive elliptical galaxies. An outstanding question remains: are the incredible infrared luminosities of sub-mm galaxies powered by huge bursts of star formation or AGN activity? To assess in detail what powers the intense IR luminosity of sub-mm galaxies it is important to obtain a complete multi-wavelength picture.

The sample of sub-mm galaxies discussed in this paper is taken from the GOODS-N SCUBA super-map (Borys et al.~2003; Pope et al.~2005). This is a purely sub-mm selected sample and it is almost completely identified at other wavelengths thanks to the deep multi-wavelength data available in the GOODS-N field (see Pope et al.~2006 for more details). 

\section{Mid-infrared photometry}

The addition of the MIPS photometry to the multi-wavelength dataset for sub-mm galaxies provides powerful constraints on the shape of the SED, the nature of the power source, and the source redshift. In the left panel of Fig.~1, we plot the $S_{24\mu m}/S_{850\mu m}$ flux ratio as a function of redshift as compared to several models. In general, this flux ratio as a function of redshift is lower than that of other ultra luminous infrared galaxies (ULIRGs) of the same luminosity at low and high redshift. This suggests that sub-mm galaxies either have higher levels of extinction or cooler dust temperatures. This can be tested further by looking in detail at the mid-IR spectrum.

\begin{figure}[th]
\begin{center}
 \plottwo{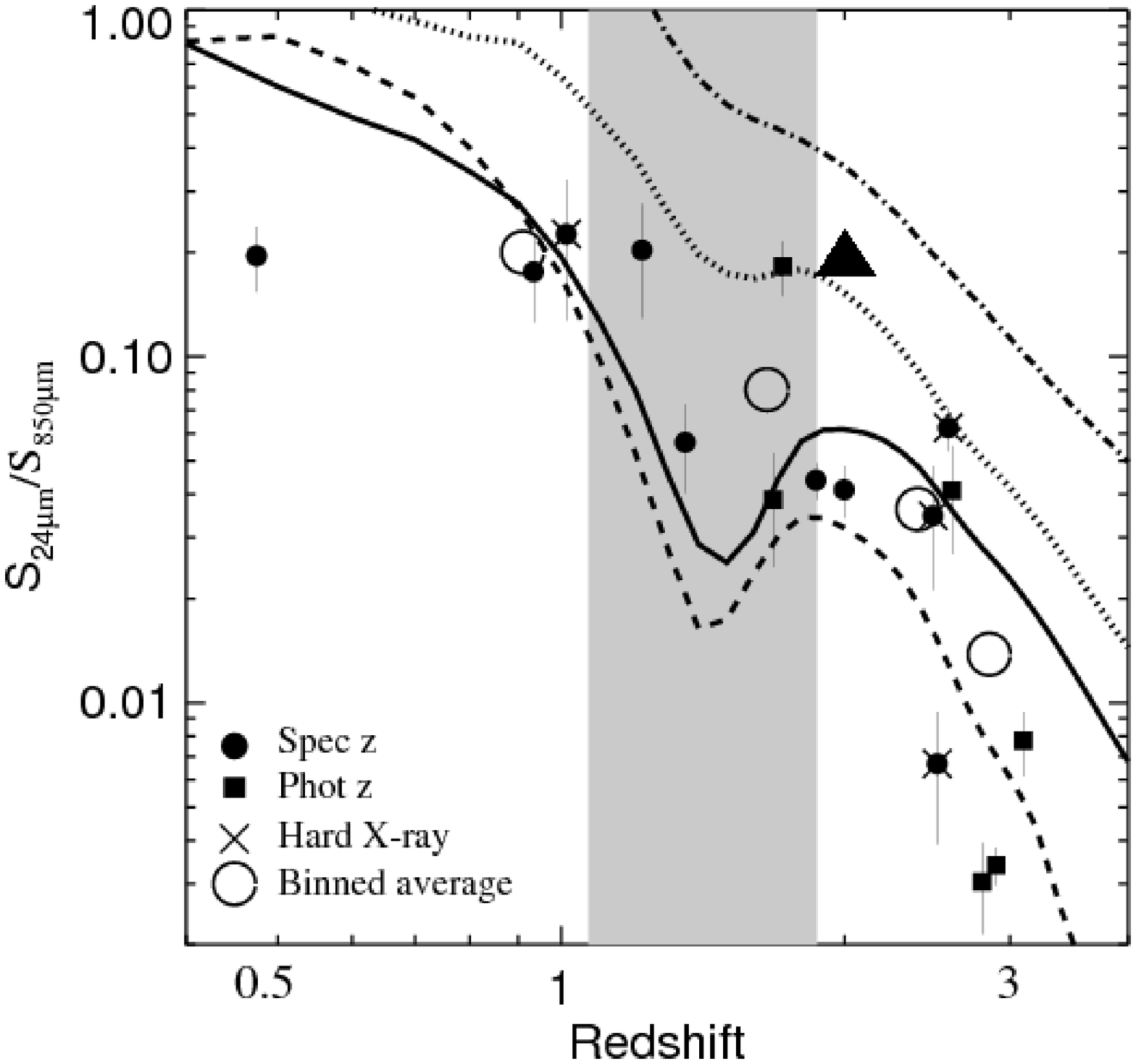}{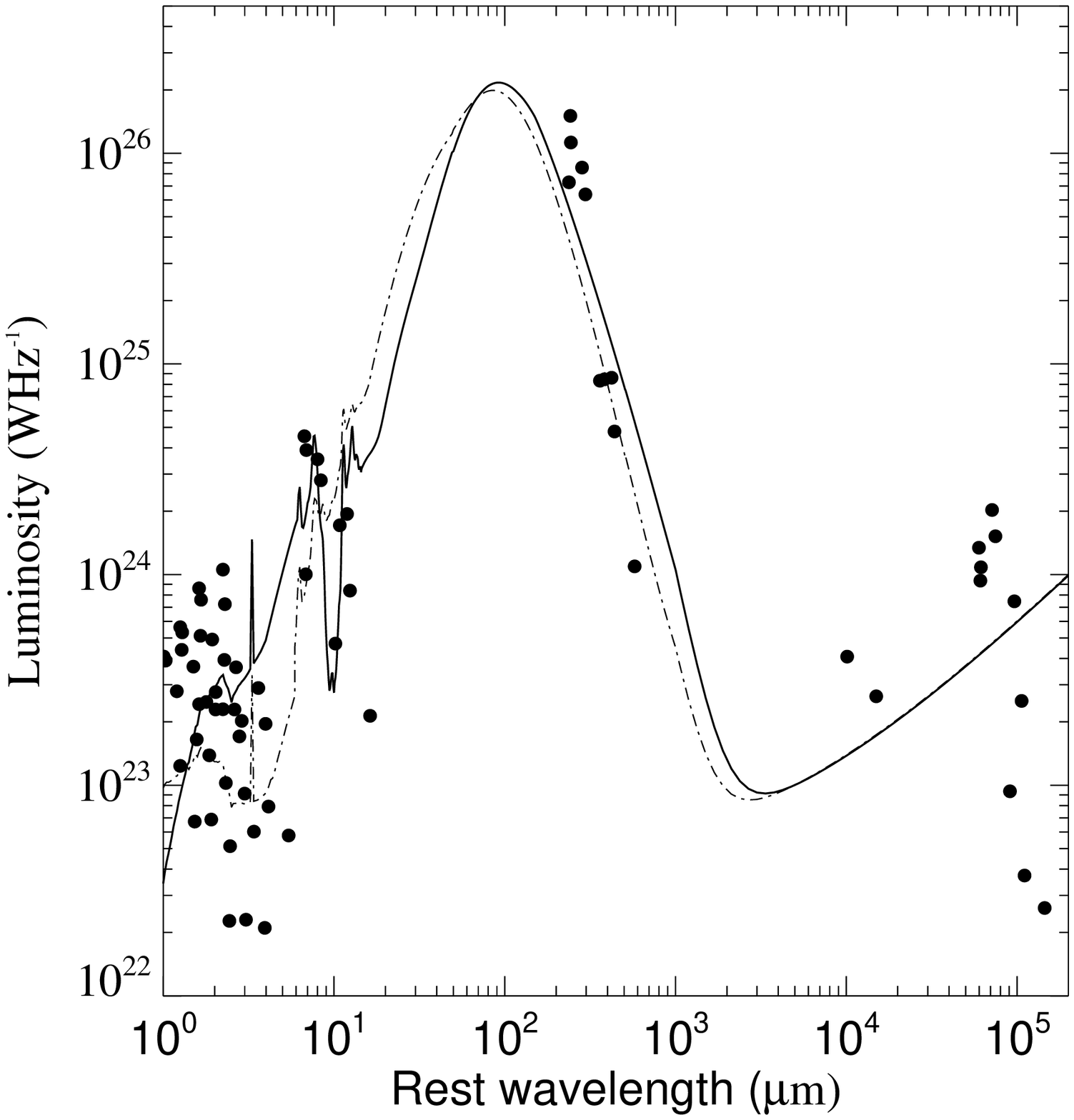}
 \caption{(left) Evolution of the $S_{24\mu m}/S_{850\mu m}$ ratio with redshift for sub-mm sources in GOODS-N as compared to several models (see Pope et al.~2006 for more details). The open circles show the average values of the sub-mm sources in four redshift bins, with roughly equal numbers of sources in each. The filled triangle is the average colour found for BzK galaxies (Daddi et al.~2005).
The dotted curve is a cool SED template from Chary \& Elbaz (2001, CE01 hereafter) models, the solid curve is the same CE01 template with additional additional extinction from Draine (2003), the dashed curve is the observed SED of Arp220 and the dash-dot curve is Mrk231, an IR-luminous AGN.
The shaded region represents the redshifts where the 9.7$\,\mu$m silicate feature passes through the 24$\,\mu$m passband. The generally lower $S_{24}/S_{850}$ ratios for the sub-mm sources suggest that they either have higher levels of extinction or cooler dust temperatures. (right) Composite rest-frame SED for GOODS-N sub-mm sources with spectroscopic redshifts. The solid curve is a modified CE01 model for a cool starburst galaxy, where we have applied additional extinction using the Draine (2003) models. The dash-dot curve is a CE01 model with the same luminosity but a warmer dust temperature.
}
\label{fig:postage}
\end{center}
\end{figure}

Using deep multi-wavelength follow-up observations of sub-mm galaxies, it is possible to study their SEDs to derive fundamental properties such as dust properties and SFRs.
In the right panel of Fig.~1, we plot a composite rest-frame SED for the sub-mm sources with spectroscopic redshift as compared to several models. We find that sub-mm galaxies have cooler average dust temperatures than those of local ULIRGs of the same luminosity (see Pope et al.~2006 for more details). This result is confirmed with the addition of deep $70\,\mu$m imaging of these sub-mm galaxies (Huynh et al.~2006). This may indicate that their far-IR emission is more extended than that of local ULIRGs, in which the majority of the IR emission comes from within the central kpc (Charmandaris et al.~2002). The SEDs of sub-mm galaxies are also different from those of their high redshift neighbours selected at near-IR wavelengths (e.g.~BzK galaxies, Daddi et al.~2005), whose mid-IR to radio SEDs are more like those of local ULIRGs. We may understand this difference in terms of the evolutionary scenario advocated by Vega et al.~(2005), in which enshrouded star forming galaxies undergo four distinct phases characterized by different mid- and far-IR colours. Under their scheme, the sub-mm galaxies, with cooler temperatures, may be an earlier phase in the star formation than BzK galaxies.

\section{Mid-infrared spectroscopy}

\begin{figure}[h]
\vspace*{-0.3in}
\begin{center}
 \plotone{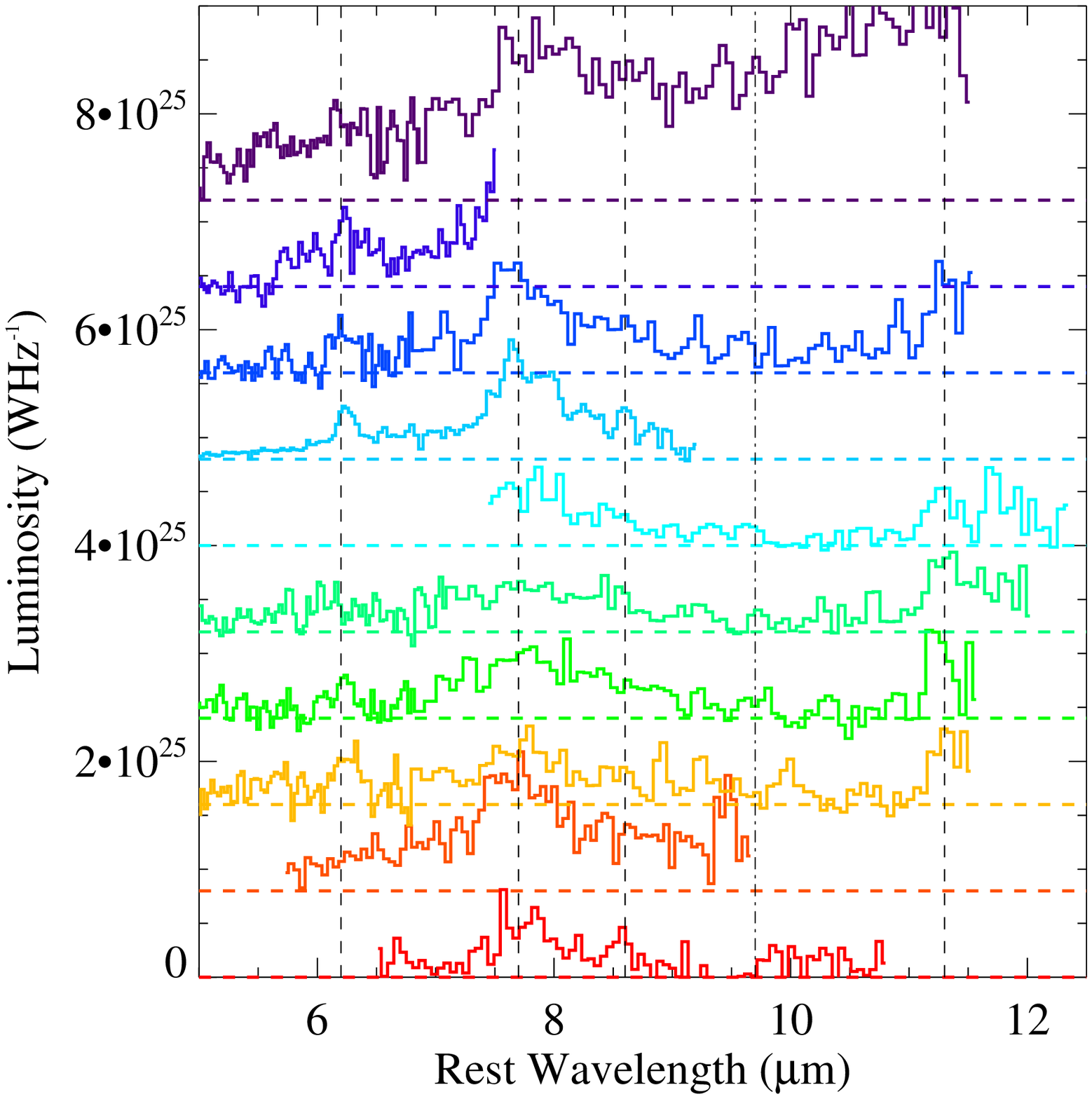}
 \caption{Raw IRS spectra of sub-mm galaxies in GOODS-N. Each spectrum has been offset for clarity and the horizontal lines show where zero is for each. Vertical lines indicate the positions of the PAH emission features and the 9.7$\,\mu$m absorption feature. 
}
\label{fig:fig3}
\end{center}
\end{figure}

Many sub-mm galaxies show evidence of an AGN, as determined either through X-ray observations (Alexander et al.~2005), or with optical spectra (Swinbank et al.~2004; Chapman et al.~2005), however it is unclear if the AGN is a significant contributor to the bolometric luminosity of the galaxy. Mid-IR spectroscopy can help determine what powers sub-mm galaxies since these wavelengths contain a number of features which are sensitive to the nature of the energy source (Clavel et al.~2000; Hudgins \& Allamandola 2004).

We obtained {\it Spitzer} IRS observations of a sample of the sub-mm galaxies and AGN in GOODS-N (GO2 PI: R.~Chary). Fig.~2 shows the rest-frame raw IRS spectra for 10 sub-mm sources from this sample, offset for clarity. All of these spectra show at least weak polycyclic aromatic hydrocarbon (PAH) emission. We have obtained several new redshifts from these IRS spectra (Pope et al.~in preparation). 
Fig.~3 shows the mean IRS spectrum for these sub-mm galaxies compared to those of several local galaxies. This composite spectrum was calculated by taking the mean of the individual spectra. Note that using the median instead of the mean does not make a significant difference to the shape of the composite spectrum. We have excluded the top source in Fig.~2 since its steeply rising mid-IR continuum is not seen in the other 9 spectra in our sample. This source is not typical of our sample and is likely to have a stronger AGN component. In the composite spectrum, we see the 6.2, 7.7, 8.6 and 11.3$\,\mu$m PAH features much more clearly. While Arp220 is often considered the typical local analog to high redshift sub-mm galaxies, it appears to have more silicate absorption and less 6.2$\,\mu$m PAH emission than the average sub-mm galaxy. Interestingly, the mid-IR spectra of the sub-mm galaxies seems to resemble that of M82 with the addition of a shallow power-law component ($\sim$$\nu^{-0.3}$). The total contribution of the power-law component to the total luminosity in this wavelength range is $\sim30\%$. M82 is a typical local starburst galaxy but it is at least 2 orders of magnitude less luminous in the IR than the average sub-mm galaxy. This composite suggests that the mid-IR emission in sub-mm galaxies is dominated by star formation activity and there is little contribution from an AGN. 
The average 7.7$\,\mu$m line to continuum ratio for this sample is $\sim2.5$--3.5. Comparing this with Figure 4 from Genzel et al.~(1998), we find that this ratio for sub-mm galaxies is consistent with the ratios for SBs and ULIRGs and not as low as those found for AGN. Our preliminary analysis suggests that sub-mm galaxies are powered primarily by SBs and not AGN activity. 

\begin{figure}[th]
\vspace{-0.3in}
\begin{center}
 \plotone{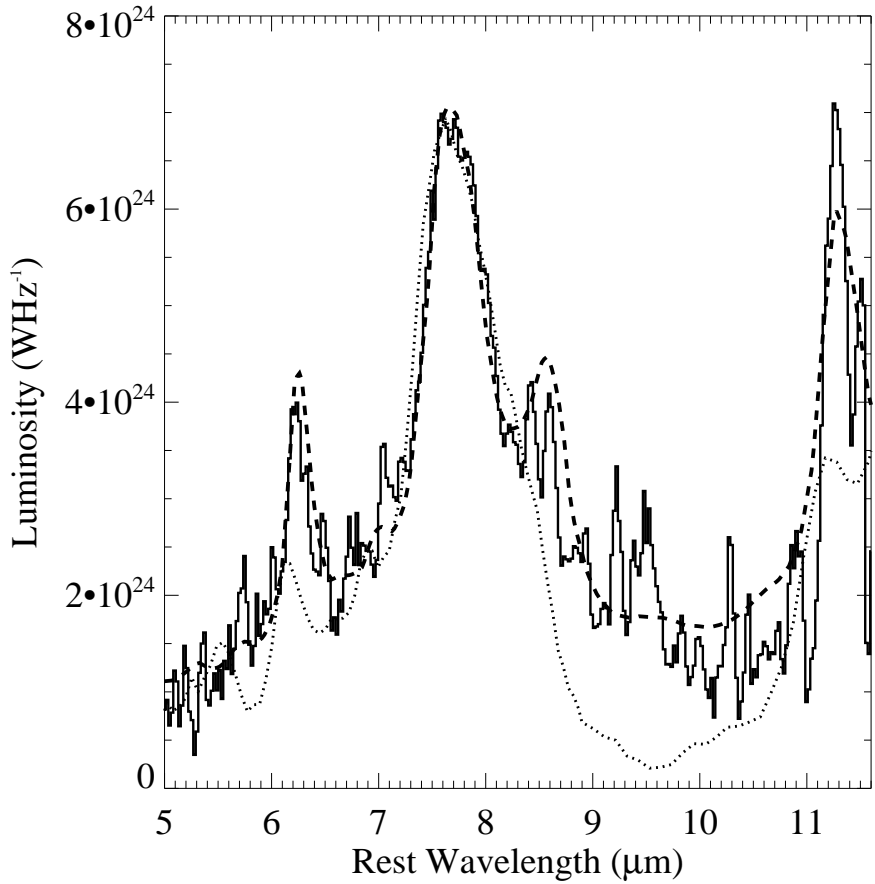}
 \caption{Composite IRS spectrum of sub-mm galaxies (Pope et al.~in preparation). The dashed curve is M82 (F{\"o}rster Schreiber et al.~2003) plus a shallow power-law component ($\sim$$\nu^{-0.3}$) while the dotted curve is Arp220 (Charmandaris et al.~1999; Fischer et al.~1999). The average sub-mm galaxy shows PAH features, some silicate absorption and a fairly flat continuum indicating little or no contribution from an AGN to the luminosity at these wavelengths.}
\label{fig:fig4}
\end{center}
\end{figure}

\section{Future Work}
We are currently working on fitting templates to quantify the level of AGN contributing to the IR luminosity and investigating various emission and absorption line strengths as diagnostics for the star formation activity.
Combined with the other multi-wavelength data that are already tabulated on these sub-mm sources, we will create a complete picture of the energetics of sub-mm galaxies (Pope et al.~in preparation). These data will help determine if sub-mm galaxies and AGN populations are connected via an evolutionary sequence.

\acknowledgements 
Thanks to the organizers of this conference for a very fruitful meeting. We also thank our collaborators involved in the GOODS-N IRS project.
This work was supported by the Natural Sciences and Engineering Research Council of Canada and the Canadian Space Agency. 
This work is based in part on observations made with the {\it Spitzer Space Telescope}, which is operated by the Jet Propulsion Laboratory, California Institute of Technology under a contract with NASA. 
Support for GOODS, part of the {\it Spitzer Space Telescope} Legacy Science Program, was provided by NASA through Contract Number 1224666 issued by JPL, Caltech, under NASA contract 1407.


\begin{thebibliography}{}
\bibitem[\protect\citeauthoryear{Alexander et al.}{2005}]{Alexander05} Alexander D.M., et al., 2005, ApJ, 632, 736
\bibitem[\protect\citeauthoryear{Barger et al.}{1998}]{1998Natur.394..248B} Barger A.J., et al., 1998, Nat, 394, 248
\bibitem[\protect\citeauthoryear{Borys et al.}{2003}]{PaperI} Borys C., et al., 2003, MNRAS, 344, 385
\bibitem[\protect\citeauthoryear{Borys et al.}{2005}]{2005ApJ...635..853B} Borys C., et al., 2005, ApJ, 635, 853
\bibitem[\protect\citeauthoryear{Chapman et al.}{2005}]{2005ApJ...622..772C} Chapman S.C., et al., 2005, ApJ, 622, 772
\bibitem[Charmandaris et al.(1999)]{1999Ap&SS.266...99C} Charmandaris, V., et al., 1999, \apss, 266, 99 
\bibitem[\protect\citeauthoryear{Charmandaris, Stacey, \& Gull}{2002}]{2002ApJ...571..282C} Charmandaris V., Stacey G.J., Gull G., 2002, ApJ, 571, 282
\bibitem[\protect\citeauthoryear{Chary \& Elbaz}{2001}]{CE01} Chary R., Elbaz D., 2001, ApJ, 556, 562
\bibitem[\protect\citeauthoryear{Clavel et al.}{2000}]{2000A&A...357..839C} Clavel J., et al., 2000, A\&A, 357, 839
\bibitem[\protect\citeauthoryear{Daddi et al.}{2005}]{2005ApJ...631L..13D} Daddi E., et al., 2005, ApJ, 631, L13
\bibitem[\protect\citeauthoryear{Draine}{2003}]{Draine} Draine B.T., 2003, ARA\&A, 41, 241
\bibitem[Fischer et al.(1999)]{1999Ap&SS.266...91F} Fischer, J., et al.\ 1999, \apss, 266, 91 
\bibitem[F{\"o}rster Schreiber et al.(2003)]{2003A&A...399..833F} F{\"o}rster Schreiber, N.~M., et al., 2003, \aap, 399, 833
\bibitem[Genzel et al.(1998)]{1998ApJ...498..579G} Genzel, R., et al.\ 
1998, \apj, 498, 579 
\bibitem[\protect\citeauthoryear{Greve et al.}{2005}]{2005MNRAS.359.1165G} Greve T.R., et al., 2005, MNRAS, 359, 1165
\bibitem[\protect\citeauthoryear{Holland et al.}{1999}]{Holland99} Holland W.S., et al., 1999, MNRAS, 303, 659
\bibitem[\protect\citeauthoryear{Hudgins \& Allamandola}{2004}]{2004ASPC..309..665H} Hudgins D.~M., Allamandola L.~J., 2004, ASPC, 309, 665 
\bibitem[\protect\citeauthoryear{Hughes et al.}{1998}]{Hughes98} Hughes D.H.~et al., 1998, Nature, 394, 241
\bibitem[\protect\citeauthoryear{Huynh et al.}{2006}]{Huynh06} Huynh M.~et al., 2006, ApJ submitted
\bibitem[\protect\citeauthoryear{Lilly et al.}{1999}]{Lilly99} Lilly S.J., et al., 1999, ApJ, 518, 641
\bibitem[\protect\citeauthoryear{Pope et al.}{2005}]{Pope05} Pope A., et al., 2005, MNRAS, 358, 149
\bibitem[\protect\citeauthoryear{Pope et al.}{2006}]{Pope06} Pope A., et al., 2006, MNRAS, 370, 1185
\bibitem[\protect\citeauthoryear{Scott et al.}{2002}]{8mJy_paperI} Scott S.E., et al., 2002, MNRAS, 331, 817
\bibitem[\protect\citeauthoryear{Smail, Ivison, \& Blain}{1997}]{Smail97} Smail I., Ivison R.J., Blain A.W., 1997, ApJ, 490, L5
\bibitem[\protect\citeauthoryear{Swinbank et al.}{2004}]{2004ApJ...617...64S} Swinbank A.M., et al., 2004, ApJ, 617, 64
\bibitem[\protect\citeauthoryear{Vega et al.}{2005}]{Vega05} Vega O., et al., 2005, MNRAS, 364, 1286 

\end{thebibliography}
\end{document}